\documentclass[aps,preprint,showpacs,superscriptaddress]{revtex4-1}  % for double-spaced preprint
\usepackage{graphicx}
\usepackage{subcaption}
\usepackage{bm}   
\usepackage{color}
\usepackage{amssymb}
\usepackage{amsmath}
\usepackage{float}
\begin{document}

\title[Rogue waves on an elliptic function background]{ Rogue waves on an elliptic function background in complex modified Korteweg-de Vries equation}

\author{N. Sinthuja}
\affiliation{Department of Nonlinear Dynamics, Bharathidasan University, Tiruchirappalli - 620 024, Tamilnadu, India}
\author{K. Manikandan}
\affiliation{Department of Nonlinear Dynamics, Bharathidasan University, Tiruchirappalli - 620 024, Tamilnadu, India}
\author{M. Senthilvelan}
\affiliation{Department of Nonlinear Dynamics, Bharathidasan University, Tiruchirappalli - 620 024, Tamilnadu, India}

\vspace{10pt}

\begin{abstract}
	With the assistance of one fold Darboux transformation formula, we derive rogue wave solutions of the complex modified Korteweg-de Vries equation on an elliptic function background. We employ an algebraic method to find the necessary squared eigenfunctions and eigenvalues. To begin we construct the elliptic function background. Then, on top of this background, we create a rogue wave.  We demonstrate the outcome for three distinct elliptic modulus values. We find that when we increase the modulus value the amplitude of rogue waves on the $dn$-periodic background decreases whereas it increases in the case of $cn$-periodic background.  
\end{abstract}

\maketitle
\section{Introduction}

A rogue wave (RW) also known as freak wave, monster wave, killer wave and gigantic wave is a short-lived and large amplitude local wave. RWs were first observed in the deep ocean and then studied both experimentally and theoretically in the field of optics, plasmonics, superfluids and capillary waves \cite{dudley,cha,kib,sch,efi,sha}.  RW is a localized wave whose height is two or three times higher than the background waves \cite{kh,wa}. During the past two decades, RWs are studied for several reasons because of its unpredictability, sudden appearance and disappearance without any warning and trace and the disastrous effects they make to the surroundings \cite{nak,ak}. The measurements and analysis of this phenomenon in ocean is extremely difficult. Hence, studies on the RWs are mainly focused either on the theoretical background or through experiments in laboratories in order to capture some information about the nature of this mysterious phenomenon.

It is found that RWs can be represented by certain rational forms of expressions that constitute solutions of the focusing nonlinear Schr\"odinger (NLS) equation which is an integrable system and plays an important role in the field of nonlinear dynamics \cite{zhaqilao,liu,ref4}. A number of mathematical methods have been used to construct RW solutions, breather solutions of various kinds and different solitary wave solutions in the nonlinear evolution equations. To name a few we cite Darboux transformation (DT) method, B\"acklund transformation, Hirota bilinear method and so on \cite{ta,ya,oh,ref1,ref2,ref3,ref5}. Among these, DT method  has been widely used to construct RW solutions of nonlinear partial differential equations.  Initially, the first order RW solution was derived by Peregrine and the second order RW solution was obtained through modified DT technique \cite{ta,wan}. Subsequently, to construct higher order RW solutions, in a simplified manner, the generalized DT was developed \cite{guo,vi}. 

\par Initially, plane wave background has been used to construct RW solutions \cite{agaf,mu,agaf1,mani,mani2}. Recently, RWs on the periodic wave background have attracted more attention because periodic oscillations are more common in the ocean surface \cite{xue}. Although the periodic wave background is considered as trigonometric functions earlier, now it has been considered in terms of Jacobian elliptic functions, namely $dn$ and $cn$. The RWs on the elliptic function wave background was initially constructed in the focusing NLS equation only through the numerical scheme \cite{kedziora}.  Subsequently, RW solutions on such a  background were derived analytically for the same NLS equation \cite{chen2}.  Progressing further, RW solutions on an elliptic function background have been attempted for the modified Kortweg-de-Vries equation (mKdV), Hirota equation, sine-Gordon equation and fifth-order NLS equation \cite{chen1,chen3,peng,pel,sinthu1}. The appearance of RW on top of a periodic wave background can be seen as a normal occurrence on the ocean surface \cite{kedziora}.  The underlying results are also applicable to the optical sciences, where fiber lines can relay pulse trains represented by periodic waves \cite{n1}. New laser techniques have been created to shape the signal amplitude in a controllable way. One can also produce Peregrine solitons with $cn$ and $dn$ background in Bose-Einstein condensates. The present study implies that it may have applications in other fields as well \cite{n3,n4}. 
\par Recently works have also been made to identify RW solutions on a double periodic background \cite{chen5,sinthu} for the NLS and Hirota equations. To obtain these kinds of solutions,  in the above cited equations, an algebraic method (nonlinearization of eigenvalue problem with Darboux transformation approach) has been adopted \cite{zhou,zhou1}. Motivated by these works, in this paper, we derive RW solution on the elliptic function background in the complex modified KdV equation by combining the DT approach with nonlinearization of eigenvalue problem. 

\par The Hirota equation is one of the widely studied extensions of the NLS equation \cite{sinthu,crabb}.   If second order dispersion and the cubic nonlinearity terms are dropped in the Hirota equation, one can obtain the complex modified KdV equation, which can also be regarded as the complex variable extension of the real valued modified KdV equation. This model is well-known as cmKdV equation in the literature and reads \cite{zhang1,he}
\begin{align}
r_t+r_{xxx}+6|r|^2r_x=0
\label{e1}
\end{align}
which possesses all the basic behaviours of an integrable model. The above equation has also been derived from physical point of view, for example as the dynamical evolution of nonlinear lattices, plasma physics, fluid dynamics, ultra short pulses in nonlinear optics and so on \cite{zhan,yuan}. 

\par The cmKdV equation admits a Lax pair of the form
\begin{subequations}
	\label{refno}
	\begin{align}
	\varphi_{x}&=M(\lambda,r) \varphi,\qquad\varphi_{t}=N(\lambda,r) \varphi, \quad M(\lambda,r)=\begin{pmatrix} \lambda & r\\ -\bar{r} & -\lambda \end{pmatrix},
	\label{e2}\\
	N(\lambda,r)&=\begin{pmatrix} -4\lambda^3-2|r|^2\lambda+r \bar{r}_x-r_x\bar{r} & -4r\lambda^2-2r_x\lambda-r_{xx}-2|r|^2r\\ 4\bar{r}\lambda^2-2\bar{r}_x\lambda+\bar{r}_{xx}+2|r|^2\bar{r} & 4\lambda^3+2|r|^2\lambda-r \bar{r}_x+r_x\bar{r} \end{pmatrix},
	\label{e3}
	\end{align}
\end{subequations}
where $\lambda$ is a spectral parameter, $r$ is the potential and the bar on top denotes its complex conjugate. Equation $(\ref{e1})$ can be obtained from the zero curvature representation $M_t-N_x+[M,N]=0$.   

\par The one-fold Darboux transformation of Eq. (\ref{e1}) is as follows \cite{he}:
\begin{align}
\hat{r}(x,t)=r(x,t)+\frac{2(\lambda_1+\bar{\lambda}_1)f_1 \bar{g}_1}{|f_1|^2+|g_1|^2},
\label{e7}
\end{align}
where $\varphi=(f_1, g_1)^T$ is a particular solution (non-zero) of the linear Eqs. (\ref{e2}) and (\ref{e3}) for a fixed eigenvalue $\lambda=\lambda_1$ and $r(x,t)$ and $\hat{r}(x,t)$ respectively represents the seed and first iterated solutions of Eq. (\ref{e1}). The linear Eqs. (\ref{e2}) and (\ref{e3}) can be solved in a variety of ways both analytically and numerically.

\par We use the one-fold Darboux transformation formula (\ref{e7}) to create RW solutions on the elliptic function wave background for the cmKdV Eq. (\ref{e1}). To begin, we determine travelling wave solutions of Eq. (\ref{e1}) in terms of elliptic functions $dn$ and $cn$. We consider an interconnection between the potential ($r$) and squared eigenfunctions ($f_1^2, \bar{g}_1^2$) of Lax system at a specific eigenvalue. From this interconnection we determine the Hamiltonian associated with the Lax pair. We note that in (\ref{refno}) the spatial part matches with the focusing NLS and Hirota equations. However, the temporal part in the spectral problem of the considered equation distincts from the NLS and Hirota equation. Upon comparing  the obtained constraints with the equations obtained through travelling wave reduction we identify the eigenvalues and the squared eigenfunctions associated for each one of the elliptic function solutions. Substituting the obtained eigenvalues, eigenfunctions and periodic travelling wave solution in the one-fold DT formula (\ref{e7}), we create the periodic background solution.  We also consider another particular solution (linearly independent from the first) of the spectral problem for the same eigenvalues in order to generate RWs on this periodic wave background. We analyze the obtained RWs on the periodic wave background by varying elliptic modulus values.

\par We organize our work as follows: In Sec. II, we construct the periodic wave background for the cmKdV Eq. (\ref{e1}).  In Sec. III, we determine the eigenvalues and squared eigenfunctions by using the method of nonlinearization of spectral problem. In Sec. IV, we construct a second solution to the spectral problem with the same eigenvalue in order to generate RW solution on the elliptic function background.  We summarize our results in Sec. 5.

\section{Periodic travelling wave solutions of cmKdV equation}
To determine a periodic wave solution of  Eq. (\ref{e1}), we assume the function $r(x,t)$ as
\begin{align}
r(x,t)= R(z) e^{i(\omega x+v_1 \omega t)},\;\;\;\;\; z=x+v_0 t,
\label{e8}
\end{align}
where $R(z)$ is a real function with $v_0$, $v_1$ and $\omega$ are the real parameters. We determine the unknown function $R(z)$ in (\ref{e8}) in terms of Jacobian elliptic functions, namely $dn$ and $cn$ by reducing Eq. (\ref{e1}) to a first order ordinary differential equation (ODE). Substituting Eq. (\ref{e8}) into Eq. (\ref{e1}) and separating real and imaginary parts we obtain the following differential equations, namely
\begin{subequations}
\begin{align}
\label{e8a} R_{zzz}+6R^2 R_z+(v_0-3\omega^{2})R_z=0, \\
\label{e9} R_{zz}+2R^3+\frac{1}{3}(v_1-\omega^{2})R=0.
\end{align}
\end{subequations}
Integrating the Eq. (\ref{e8a}) with respect to $z$, we obtain 
\begin{align}
\label{e9a} R_{zz}+2R^3+(v_0-3\omega^{2})R=0,
\end{align}
where we have taken the integration constant as zero.  For our convenience, by imposing the constraint $\frac{1}{3}(v_1-\omega^{2})=(v_0-3\omega^{2})=-4a$ in the above two Eqs. (\ref{e9}) and (\ref{e9a}), respectively, we observe that the above two equations reduce to a single equation as
\begin{align}
R_{zz}+2R^3-4a R=0,
\label{e121}
\end{align}
where $a$ is a real parameter.  Integrating Eq. (\ref{e121}) one more time with respect to $z$ we get the following first order ODE, that is
\begin{align}
R^2_z+w(R)=0,\quad w(R)=R^4-4a R^2-8d,
\label{e11}
\end{align}
where $d$ is the real constant. We note here that Eq. (\ref{e11}) has four roots, which are symmetric and they can be represented as $\pm R_1$ and $\pm R_2$.  We obtain elliptic functions $dn$- and $cn$- solutions for two different choices, namely (i) if all four roots are real and (ii) two roots are real and two roots are complex conjugates, respectively.  In the first case (all roots are real) we can arrange the roots as $-R_1\leq -R_2\leq R_2\leq R_1$. The first order equation $(\ref{e11})$ has solutions in either $[-R_1,-R_2]$ or $[R_2,~ R_1]$. The transformation ($R\rightarrow-R$) connects the two solutions, so the positive solution in $[R_2,~ R_1]$ can be considered without losing generality and this solution can be represented by  
\begin{align}
&R(z)=R_1~dn[R_1~z, k] = R_1~dn[R_1(x+v_0 ~t), k] , \quad k^2=\frac{R_1^2-R_2^2}{R_1^2}.
\label{e12}
\end{align}
In the second case, we consider two roots of Eq. (\ref{e11}) are real $(\pm R_1)$ and the other two roots are complex $(\pm R_2=\pm i T_2)$.  So the solution in $[-R_1, R_1]$ can be represented as 
\begin{align}
&R(z)=R_1~cn[\beta~z, k] = R_1~ cn[\beta(x+v_0~ t),k],\quad \beta^2=R_1^2+T_2^2,\quad k^2=\frac{R_1^2}{R_1^2+T_2^2},
\label{e13}
\end{align}
where $k\in(0,1)$ is the elliptic modulus. The expressions which connect the roots $\pm R_1$ and $\pm R_2$ with the parameters ($a$ and $d$)  in Eq. (\ref{e11}) are given by 
\begin{align}
a=\frac{1}{4}(R_1^2+R_2^2),\quad d=-\frac{1}{8}R_1^2 R_2^2.
\label{n2}
\end{align}

\begin{figure}[!ht]
		\begin{subfigure}{0.45\linewidth}
		\centering
			\caption{}
			\includegraphics[width=\linewidth]{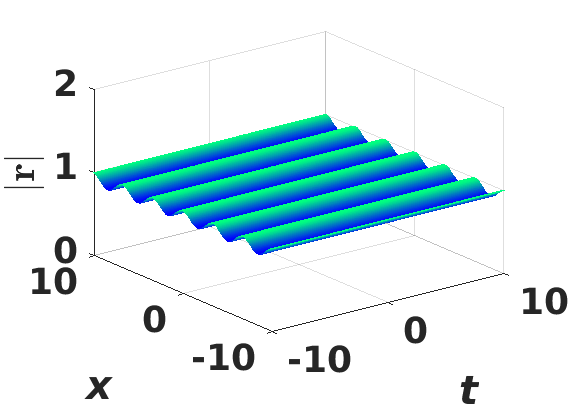}
			%\caption{}
		\end{subfigure}
		\begin{subfigure}{0.45\linewidth}
		\centering
		\caption{}
			\includegraphics[width=\linewidth]{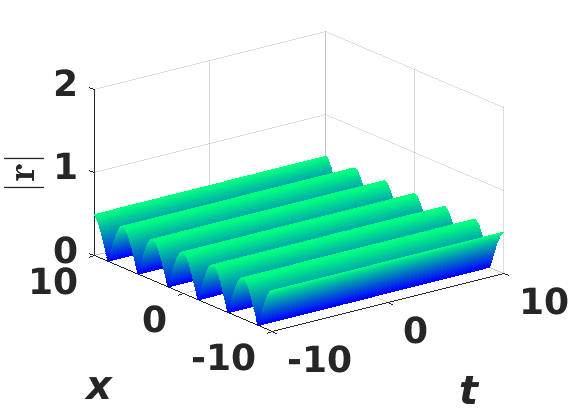}
			%\caption{}
		\end{subfigure}
		\begin{subfigure}{0.45\linewidth}
		\centering
			\caption{}
			\includegraphics[width=\linewidth]{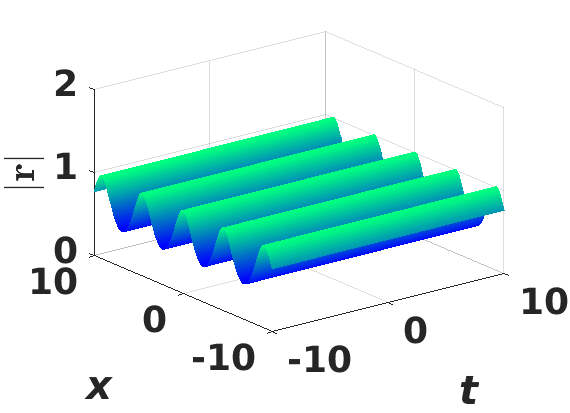}
			%\caption{}
		\end{subfigure}
		\begin{subfigure}{0.45\linewidth}
		\centering
			\caption{}
			\includegraphics[width=\linewidth]{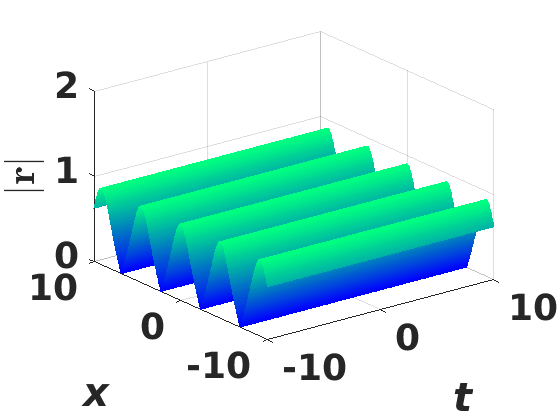}
			%\caption{}
		\end{subfigure}
	%\end{center}
	\vspace{-0.1cm}
	\caption{Periodic wave profile of (\ref{e8}) with (\ref{e12}) and (\ref{e13}).  (a)  $dn$- periodic wave background and (b) $cn$- periodic wave background  for $k=0.5$ and (c)  $dn$- periodic wave background and (d) $cn$- periodic wave background  for $k=0.9$. $\omega=0.05$ for (a) and (c) and $\omega=0.01$ for (b) and (d). }
	\label{dnnfig1}
\end{figure}
\par Figures \ref{dnnfig1}(a)-(b) represent the periodic wave pattern of Eqs. (\ref{e12}) and (\ref{e13}) using (\ref{e8}) for $k=0.5$. When we increase the $k$ value from $0.5$ to $0.9$, we observe that the amplitudes of both the periodic waves ($dn,~cn$) increase as seen in Figs. \ref{dnnfig1}(c)-(d).
\section{Eigenvalues and Eigenfunctions of Eq. (\ref{refno})}
In the DT method, conventionally, we assume a seed solution and substituting this solution in the Lax pair equations and solving them consistently, we obtain the eigenfunctions and eigenvalues. This conventional procedure is very difficult to follow in the present case with elliptic functions as seed solution. To overcome this obstacle and to determine the admissible eigenvalues and eigenfunctions, we adopt the nonlinearization of Lax pair procedure. In this procedure, we interrelate the potential with squared eigenfunctions and derive two first- order invariants on the potential $r(x,t)$ by appropriately using the Lax pair Eqs. (\ref{e2}) and (\ref{e3}).  Using this invariants we rewrite the polynomial $(Q(\lambda))$ in terms of real parameters ($a, v_0, v_1$ and $d$). We then determine the admissible eigenvalues ($(\lambda_1)$, $(\lambda_2)$) and the corresponding squared eigenfunctions $(f^2_1, ~\bar{g}^2_1,~f_1g_1)$.

We consider the following constraint, namely
\begin{align}
r=f^2_1+\bar{g}^2_1,
\label{e14}
\end{align}
where $(f_1,g_1)^T$ is a non-zero solution of the Lax pairs (\ref{e2}) and (\ref{e3}) at $\lambda=\lambda_1$.  Differentiating Eq. (\ref{e14}) with repect to $x$, we obtain the following differential expressions, namely
\begin{align}
r_x+2irA_0-2(\lambda_1 f^2_1-\bar{\lambda}_1 \bar{g}^2_1)&=0,
\label{e20}\\
r_{xx}+2iv_0r_x+2|r|^2r-4v_1r&=0,
\label{e211}
\end{align} 
where $v_0=A_0+i(\lambda_1-\bar{\lambda}_1)$ and $v_1=H_0+iA_0(\lambda_1-\bar{\lambda}_1)+|\lambda_1^2|$ and $A_0=i(f_1g_1-\bar{f}_1\bar{g}_1)$, 
$A_1=\lambda_1 f_1g_1+\bar{\lambda}_1\bar{f}_1 \bar{g}_1+\frac{1}{2}(|f_1|^2+|g_1|^2)^2$ are the constants of motion and $H_0$ is the Hamiltonian of the system which is given by
\begin{align}
\label{eh1}
H_0=A_1-\frac{1}{2}A^2_0=\lambda_1 f_1g_1+\bar{\lambda}_1\bar{f}_1 \bar{g}_1+\frac{1}{2}(f^2_1+\bar{g}^2_1)(\bar{f}^2_1+g^2_1).
\end{align} 

The second-order differential Eq. (\ref{e211}) admits the following Lax representation, that is
\begin{align}
\frac{d}{dx}J(\lambda)=[U(\lambda,r),J(\lambda)],\qquad \lambda\in \mathbb{C},
\label{ee1}
\end{align}
where the matrices $U(\lambda,r )$ and $J(\lambda)$ is given by
\begin{align}
%\label{ee21}
U(\lambda,r)=\begin{bmatrix}
\lambda & f^2_1+\bar{g}^2_1  \\
-(\bar{f}^2_1+g^2_1)& -\lambda
\end{bmatrix},\nonumber
\end{align}
\begin{align}
\label{ee2}
J(\lambda)=\begin{bmatrix}
1- \left(\frac{f_{1} g_{1}}{\lambda-{\lambda}_1}-\frac{\bar{f}_{1} \bar{g}_{1}}{\lambda+{\bar{\lambda}}_1}\right) & \frac{f^2_{1} }{\lambda-{\lambda}_1}+\frac{ \bar{g}^2_{1}}{\lambda+{\bar{\lambda}}_1}  \\
-\frac{f^2_{1} }{\lambda-{\lambda}_1}-\frac{ \bar{g}^2_{1}}{\lambda+{\bar{\lambda}}_1}& -1+ \left(\frac{f_{1} g_{1}}{\lambda-{\lambda}_1}-\frac{\bar{f}_{1} \bar{g}_{1}}{\lambda+{\bar{\lambda}}_1}\right)
\end{bmatrix}.
\end{align}

One can get Eq. (\ref{e211}) by considering the upper right component that appear in the Lax equation $(\ref{ee1})$ with $(\ref{ee2})$.
To get the first order invariants of (\ref{e211}), we evaluate the determinant of $J(\lambda)$.  Doing so, we obtain
\begin{align}
\text{det}J(\lambda)=&-\left[\frac{(\lambda-\lambda_1)(\lambda+\bar{\lambda}_1)+i A_0(\lambda-\lambda_1+\bar{\lambda}_1)+\frac{1}{2}(A^2_0+|r|^2)-A_1}{(\lambda-\lambda_1)(\lambda+\bar{\lambda}_1)}\right]^2\nonumber\\
&+\frac{[(\lambda-\lambda_1+\bar{\lambda}_1+iA_0)r+\frac{1}{2}r_x][(\lambda-\lambda_1+\bar{\lambda}_1+iA_0)\bar{r}-\frac{1}{2}\bar{r}_x]}{(\lambda-\lambda_1)^2(\lambda+\bar{\lambda}_1)^2}.
\label{Ee1}
\end{align}

Equation (\ref{Ee1}) can be rewritten in the following form
\begin{align}
\text{det}J(\lambda)= -\frac{Q(\lambda)}{(\lambda-\lambda_1)^2(\lambda+\bar{\lambda}_1)^2},
\label{n3}
\end{align}
where $Q(\lambda)$ is described by
\begin{align}
Q(\lambda)=(\lambda^2+iv_0\lambda+\frac{1}{2}|r^2|-v_1)^2-(r\lambda+\frac{1}{2}r_x+iv_0r)(\bar{r}\lambda-\frac{1}{2}\bar{r}_x+iv_0\bar{r}).
\label{n4}
\end{align}
In order to construct $Q(\lambda)$ which is independent of $(x,t)$ we expand the polynomial in powers of $\lambda$.  The resultant action gives the following two differential constraints for the second order equation (\ref{e211}), namely
\begin{align}
i(\bar{r}r_x-r\bar{r}_x)-2v_0|r|^2-4a_1=0,
\label{e22}\\
|r_x|^2+|r|^4-4v_1|r|^2-8d=0.
\label{e23}
\end{align}

The differential constraint (\ref{e23}) resembles with the first-order equation (\ref{e11}) obtained through travelling wave reduction (with some differences in the coefficients).  One may also notice that the eigenvalues $(\lambda_1)$ do not appear explicitly in Eqs. (\ref{e12}) and (\ref{e13}). After implementing the nonlinearization of Lax pair procedure, we obtain the coefficients ($a$  and $d$) with the presence of eigenvalues $(\lambda_1)$ for both the periodic waves.
By using the above constraints ((\ref{e22}) and (\ref{e23})), we can rewrite the polynomial $Q(\lambda)$ in Eq. (\ref{n4}) in terms of these parameters as follows:
\begin{align}
Q(\lambda)=\lambda^4+2iv_0\lambda^3-(c^2+2v_1)\lambda^2+2i(a_1-v_1v_0)\lambda+v_1^2-2a_1v_0+2d.
\label{n5}
\end{align}

 Let us consider $v_0=0$ and $a_1=0$, then the polynomial $Q(\lambda)$ becomes
 \begin{align}
 Q(\lambda)=\lambda^4-\frac{1}{2}(R_1^2+R_2^2)\lambda^2+\frac{1}{16}(R_1^2-R_2^2)^2.
 \label{n6}
 \end{align}  
 The admissible two pairs of eigenvalues are
 \begin{align}
 \lambda_1=\pm\frac{R_1+R_2}{2},\quad \lambda_2=\pm\frac{R_1-R_2}{2},
 \end{align}
 in which $R_1$ is real and $R_2$ can act as a real or purely imaginary one with respect to Eqs. (\ref{e12}) and (\ref{e13}).

The explicit expressions of the eigenvalues $(\lambda_1)$ and eigenfunctions $(f^2_1,~\bar{g}^2_1,~f_1g_1)$ of the $dn$- periodic wave with $R_1=1$, $R_2=\sqrt{1-k^2}$ and $v_0=0$ are given by 
\begin{subequations}
	\label{s1}
	\begin{align}
	& \text{dn periodic wave}~~~~~ : R(x)=dn(x,k),\quad k\in (0,1),
	\label{n7}\\
	&\text{Eigenvalues}~~~~~~~~~~~~~: \lambda_{1}=\pm\frac{1}{2}(1+ \sqrt{1-k^2}),\quad\lambda_{2}=\pm\frac{1}{2}(1- \sqrt{1-k^2}),
	\label{e28}\\
	&\text{Eigenfunctions}~~~~~~~~: f^2_1=\frac{2\lambda_1 r+r_x}{2(\lambda_1+\bar{\lambda}_1)},\quad \bar{g}^2_1=\frac{2\bar{\lambda}_1 r-r_x}{2(\lambda_1+\bar{\lambda}_1)}, \nonumber \\
	&\qquad	\qquad \quad \quad\quad\quad ~~~ f_1g_1=-\frac{1}{4\lambda_1}[|r(x,t)|^2+\sqrt{1-k^2}].
	\label{e31}
	\end{align}
\end{subequations}
 Similarly, the explicit expressions of the eigenvalues $(\lambda_1)$ and eigenfunctions $(f^2_1,~\bar{g}^2_1,~f_1g_1)$ of the $cn$- periodic wave with $R_1=k$, $R_2=iT_2=i \sqrt{1-k^2}$ and $v_0=0$ read
\begin{subequations}
	\label{s2}
	\begin{align}
	& \text{cn periodic wave}~~~~~~: R(x)=k~cn(x,k),\quad k\in (0,1),
	\label{n8}\\
	&\text{Eigenvalues}~~~~~~~~~~~~~~: \lambda_{1}=\pm\frac{1}{2}(k+ i\sqrt{1-k^2}),\quad \lambda_{2}=\pm\frac{1}{2}(k- i\sqrt{1-k^2}),
	\label{e29}\\
	&\text{Eigenfunctions}~~~~~~~~~: f^2_1=\frac{2\lambda_1 r+r_x}{2(\lambda_1+\bar{\lambda}_1)},\quad \bar{g}^2_1=\frac{2\bar{\lambda}_1 r-r_x}{2(\lambda_1+\bar{\lambda}_1)},	\nonumber \\
	&\qquad	\qquad \quad \quad\quad\quad~~~~	f_1g_1=-\frac{1}{2k}[|r(x,t)|^2+ik\sqrt{1-k^2}].
	\label{e32}
	\end{align}
\end{subequations}
The obtained expressions for the squared eigenfunctions ($f^2_1$ and $\bar{g}^2_1$), namely (\ref{e31}) and (\ref{e32}) satisfy the constraint (\ref{e14}).  We recall here that since $f_1$ and $g_1$ satisfies the Hamiltonian system (\ref{eh1}) with the constraint (\ref{e14}), the function $r$ solves Eq. (\ref{e1}) (for the proof one may refer \cite{zhou}).

We substitute the obtained eigenvalues $(\lambda_1)$, periodic eigenfunctions $(f_1^2,\bar{g}^2_1,f_1g_1)$ and periodic wave solutions $r(x,t)$ in the one-fold DT formula $(\ref{e7})$. The resultant action generates a periodic wave background.  On the surface of these two periodic waves now we create a RW.   Hence, we proceed to construct a second solution to the spectral problem (\ref{refno}) for the same eigenvalue $\lambda=\lambda_1$ and obtain the desired RW solution on the elliptic function background.
\begin{figure}[!ht]
	\begin{center}
		\begin{subfigure}{0.45\textwidth}
			\caption{}
			\includegraphics[width=\linewidth]{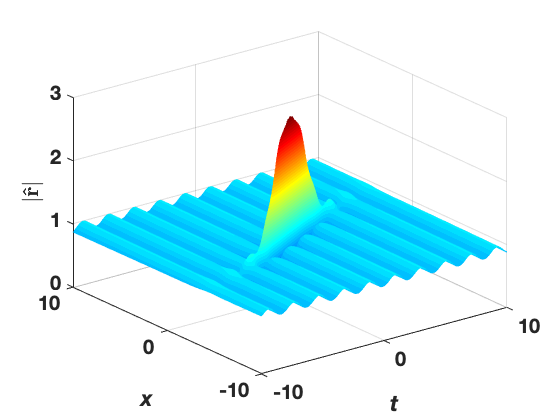}
			%\caption{}
		\end{subfigure}
		\begin{subfigure}{0.45\textwidth}
			\caption{}
			\includegraphics[width=\linewidth]{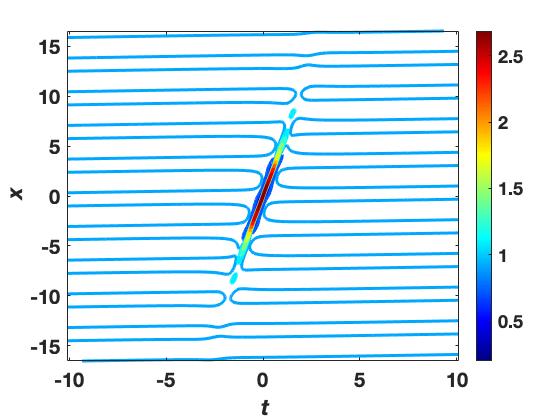}
			%\caption{}
		\end{subfigure}
		\begin{subfigure}{0.45\textwidth}
			\caption{}
			\includegraphics[width=\linewidth]{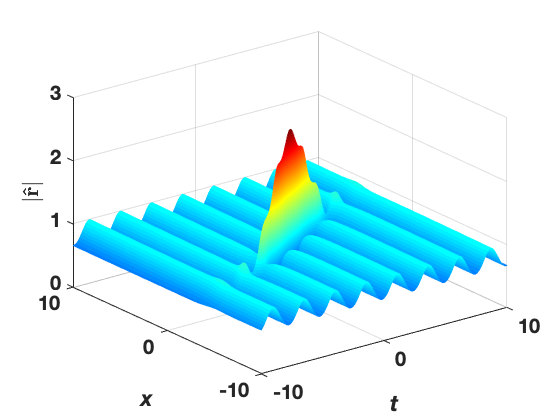}
			%\caption{}
		\end{subfigure}
		\begin{subfigure}{0.45\textwidth}
			\caption{}
			\includegraphics[width=\linewidth]{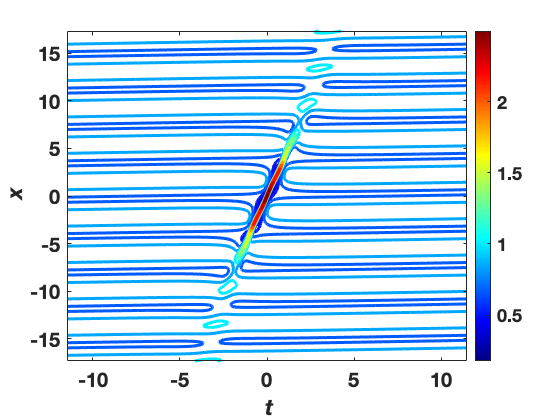}
			%\caption{}
		\end{subfigure}
		\begin{subfigure}{0.45\textwidth}
			\caption{}
			\includegraphics[width=\linewidth]{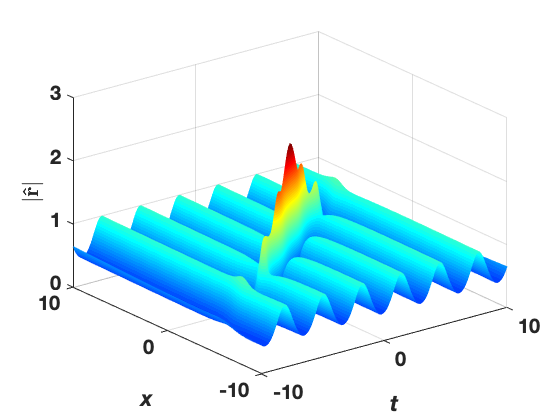}
			%\caption{}
		\end{subfigure}
		\begin{subfigure}{0.45\textwidth}
			\caption{}
			\includegraphics[width=\linewidth]{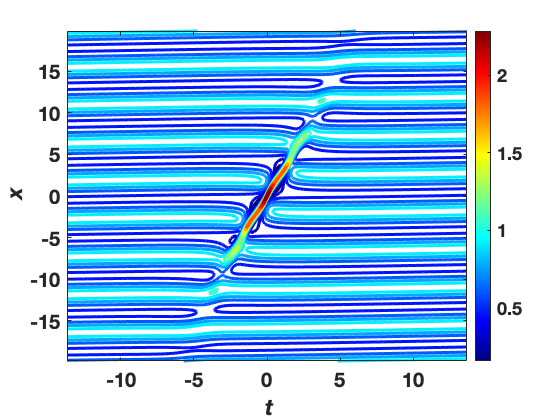}
			%\caption{}
		\end{subfigure}
	\end{center}
	\vspace{-0.3cm}
	\caption{RWs on the dn-periodic background of (\ref{e38}) with (\ref{e12}).  (a)-(b) $k=0.5$, (c)-(d) $k=0.75$ and (e)-(f) $k=0.9$ with $\omega=0.05$; Panels (b), (d) and (f) are corresponding contour plots of (a), (c) and (e).}
	\label{dnnfig2}
\end{figure}

\section{Non-periodic solutions of Lax pair}
 We construct a non-periodic solution which is in rational form for the Eqs.  (\ref{e2}) and (\ref{e3}) with the same eigenvalues. The non-periodic second solution which we intend to construct should grow linearly in $x$ and $t$ and should act as RWs.  We create the desired form of solution with the following expressions, that is
\begin{align}
\label{e33}
\hat{f}_1 = f_1 \delta_1-\frac{2\bar{g}_1}{|f_1|^2+|g_1|^2}, \quad \hat{g}_1 =  g_1 \delta_1+\frac{2\bar{f}_1}{|f_1|^2+|g_1|^2},
\end{align}
with $\delta_1$ is of the form
\begin{align}
\label{e37}
\delta_1(x,t)= & \int_{x_0}^{x} W_1(x',t)dx'+ \int_{t_0}^t W_2(x_0,t')dt',
\end{align}
where
\begin{subequations}
	\label{cm}
	\begin{align}
	\label{e35}
	W_1&=\frac{\partial \delta_1}{\partial x}:=  -\frac{4(\lambda_1+\bar{\lambda_1})\bar{f_1}\bar{g}_1}{\left(|f_1|^2+|g_1|^2\right)^2},\\
	\label{e36}
	W_2&=\frac{\partial \delta_1}{\partial t}: =-\frac{2(S_1 \bar{f}^2_1-S_2\bar{g}^2_1-S_3\bar{f}_1\bar{g}_1)}{(|f_1|^2+|g_1|^2)^2},
	\end{align}
\end{subequations}
with
$S_1=r_{xx}+2|r|^2r+2\lambda_1(r_x+2\lambda_1)-2\bar{\lambda}_1(|r|^2+2\bar{\lambda}^2_1),~S_2=2(\lambda_1+\bar{\lambda}_1)\bar{r}_x$ and $S_3=r_{xx}+2|r|^2r+4\lambda_1(|r|^2+2\lambda^2_1)-2\bar{\lambda}_1(r_x-2\bar{\lambda}^2_1-|r|^2-2r\bar{\lambda}_1)$ and $(x_0,t_0)$ is arbitrarily fixed. The above expressions $W_1$ and $W_2$ are determined by inserting Eq. (\ref{e33}) into Eqs. (\ref{e2}) and (\ref{e3}) and utilizing the later equations for $\varphi =(\hat{f}_1,\hat{g}_1)^T$.  Note that the second solution is denoted by the letters $\hat{f}_1$ and $\hat{g}_1$.
\begin{figure}[ht]
	\begin{subfigure}[c][1\width]{0.32\textwidth}
		\centering
		\caption{}
		\includegraphics[width=\linewidth]{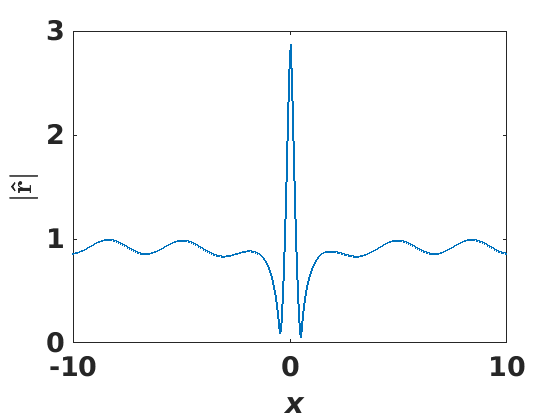}
	\end{subfigure}
	\hfill 	
	\begin{subfigure}[c][1\width]{0.32\textwidth}
		\centering
		\caption{}
		\includegraphics[width=\linewidth]{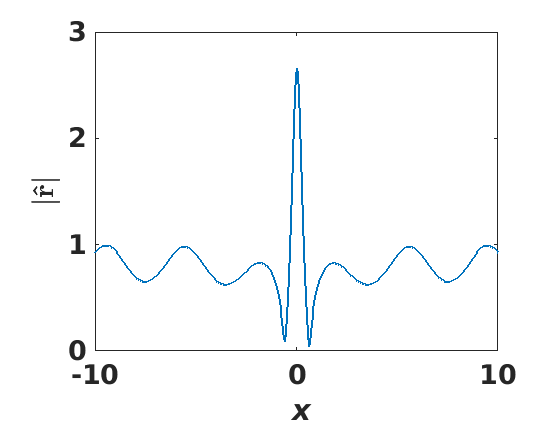}
	\end{subfigure}
	\hfill	
	\begin{subfigure}[c][1\width]{0.32\textwidth}
		\centering
		\caption{}
		\includegraphics[width=\linewidth]{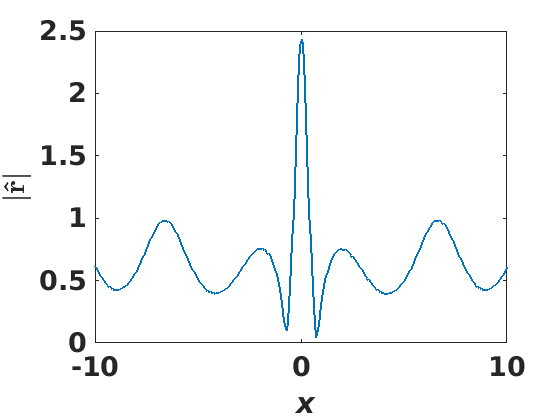}
	\end{subfigure}
	\vspace{-0.3cm}
	\caption{Two dimensional plot for RWs on the dn-periodic background of (\ref{e38}) with (\ref{e12}) at $t=0$.  (a) $k=0.5$, (b) $k=0.75$ and (c) $k=0.9$ with $\omega=0.05$.}
	\label{dnnfig3}
\end{figure}

In Eq. (\ref{e7}) we substitute the obtained periodic solutions $r(x,t)$ (\ref{e8}) along with (\ref{e12}) and (\ref{e13}) and also the eigenfunctions of second solution $\varphi =(\hat{f_1},\hat{g}_1)^T$ of the linear Eqs. (\ref{e2})-(\ref{e3}) with $\lambda=\lambda_1$.  The resultant action yields a new solution to the cmKdV Eq. (\ref{e1}) of the form
\begin{align}
\label{e38}
\hat{r}(x,t)= & r(x,t)+\frac{2(\lambda_1+\bar{\lambda_1})\hat{f_1}\bar{\hat{g}}_1}{|\hat{f_1}|^2+|\hat{g}_1|^2},  
\end{align}
where $\hat{f_1}$ and $\hat{g}_1$ are given in (\ref{e33}).

\begin{figure}[!ht]
	\begin{center}
		\begin{subfigure}{0.45\textwidth}
			\caption{}
			\includegraphics[width=\linewidth]{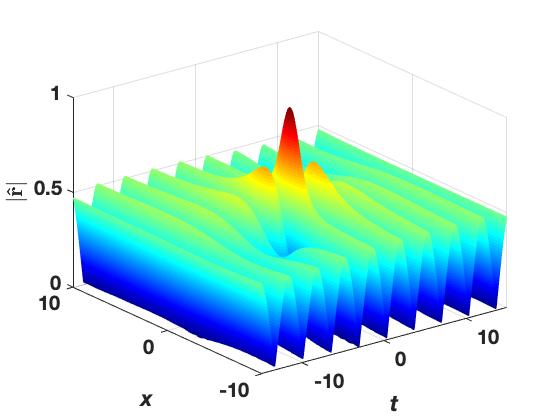}
			%\caption{}
		\end{subfigure}
		\begin{subfigure}{0.45\textwidth}
			\caption{}
			\includegraphics[width=\linewidth]{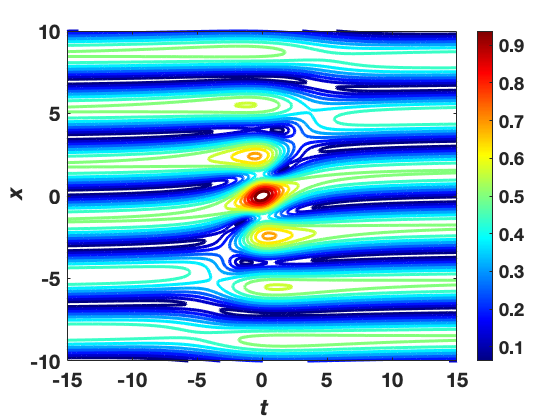}
			%\caption{}
		\end{subfigure}
		\begin{subfigure}{0.45\textwidth}
			\caption{}
			\includegraphics[width=\linewidth]{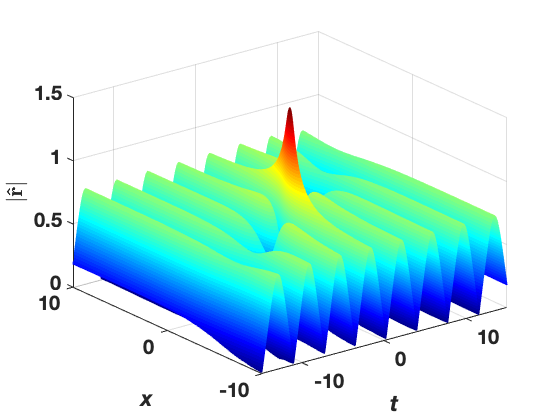}
			%\caption{}
		\end{subfigure}
		\begin{subfigure}{0.45\textwidth}
			\caption{}
			\includegraphics[width=\linewidth]{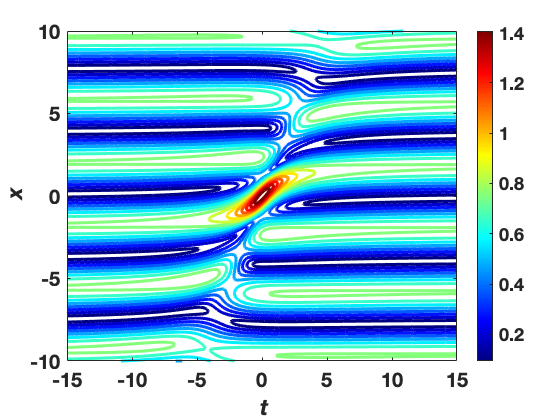}
			%\caption{}
		\end{subfigure}
		\begin{subfigure}{0.45\textwidth}
			\caption{}
			\includegraphics[width=\linewidth]{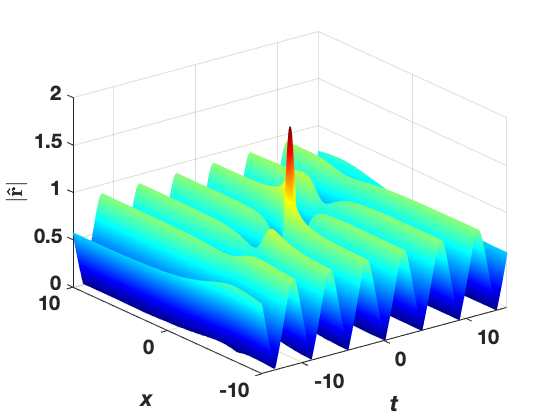}
			%\caption{}
		\end{subfigure}
		\begin{subfigure}{0.45\textwidth}
			\caption{}
			\includegraphics[width=\linewidth]{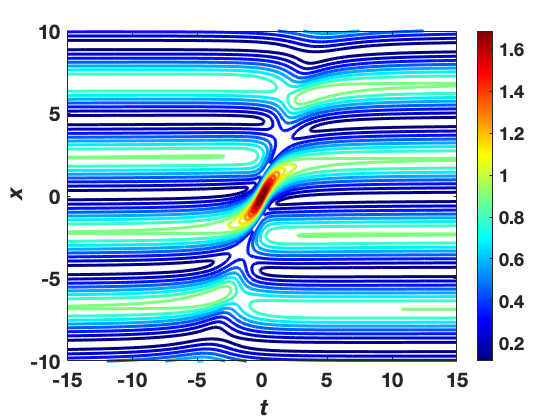}
			%\caption{}
		\end{subfigure}
	\end{center}
	\vspace{-0.3cm}
	\caption{RWs on the cn-periodic background of (\ref{e38}) with (\ref{e13}). (a)-(b) $k=0.5$, (c)-(d) $k=0.75$ and (e)-(f) $k=0.9$ with $\omega=0.01$; Panels (b), (d) and (f) are corresponding contour plots of (a), (c) and (e).}
	\label{dnnfig5}
\end{figure}
\begin{figure}[ht]
	\begin{subfigure}[c][1\width]{0.32\textwidth}
		\centering
		\caption{}
		\includegraphics[width=\linewidth]{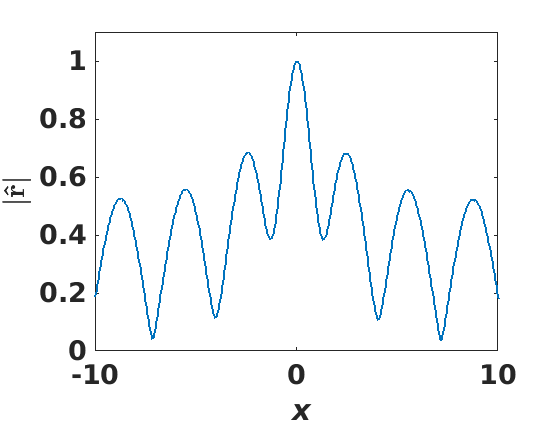}
	\end{subfigure}
	\hfill 	
	\begin{subfigure}[c][1\width]{0.32\textwidth}
		\centering
		\caption{}
		\includegraphics[width=\linewidth]{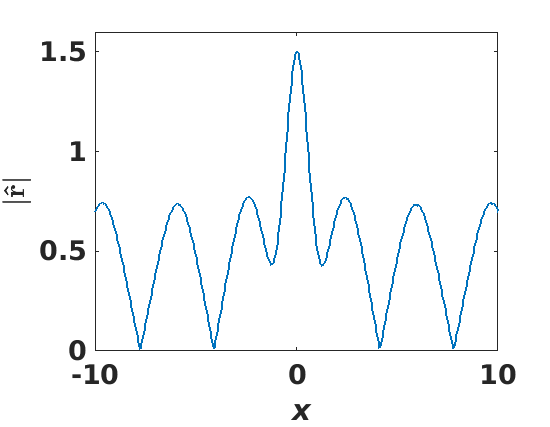}
	\end{subfigure}
	\hfill	
	\begin{subfigure}[c][1\width]{0.32\textwidth}
		\centering
		\caption{}
		\includegraphics[width=\linewidth]{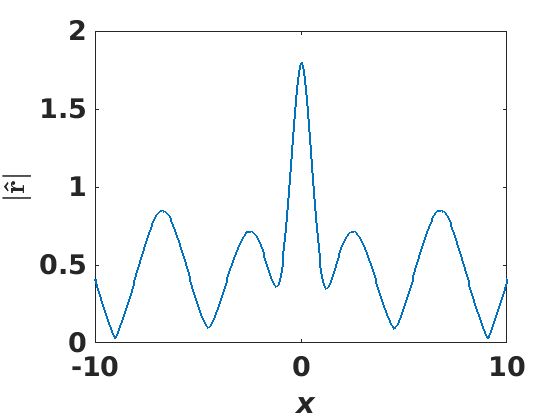}
	\end{subfigure}
	\vspace{-0.3cm}
	\caption{Two dimensional plot for RWs on the cn-periodic background of (\ref{e38}) with (\ref{e13}) at $t=0$.  (a) $k=0.5$, (b) $k=0.75$ and (c) $k=0.9$ with $\omega=0.01$.}
	\label{dnnfig4}
\end{figure}

If we consider the seed solution $r(x,t)$ in $dn$ periodic wave form with $\lambda_1=\frac{1}{2}[1+\sqrt{1-k^2}]$, the new solution reveals RW on the $dn$-elliptic function background. Similarly, if we take the seed solution as $cn$ periodic wave form with $\lambda_1=\frac{1}{2}[k+i\sqrt{1-k^2}]$, the new solution turns out to be the RW solution on the $cn$-elliptic function background.

The surface plots of $|\hat{r}|$ for three different values of the elliptic modulus are illustrated in Fig. \ref{dnnfig2}. In this plot, we represent RW that arises on the $dn$-periodic wave background using Eqs. (\ref{e38}) and (\ref{e8}) with (\ref{e12}). In Fig. \ref{dnnfig2}(a), we present the RW on the periodic background for $k=0.5$. We notice that RW attains its maximum amplitude ($|\hat{r}|=2.866$) at its origin, $(x_0,t_0)$. The corresponding contour plot is illustrated in  Fig.~\ref{dnnfig2}(b).  Figures \ref{dnnfig2}(c) and \ref{dnnfig2}(e) represent RWs on such background for two different values of elliptic modulus $(k=0.75,~k=0.9)$ and the corresponding contour plots are displayed in Figs. \ref{dnnfig2}(d) and \ref{dnnfig2}(f), respectively. The amplitude of the RWs is different for each $k$ value and it is found to be $|\hat{r}|=2.661,~2.436$ for $k=0.75$ and $k=0.9$, respectively. As the $k$ value is increased, we observe that the frequency of the periodic wave changes.  To have a better understanding on the solutions we also present two-dimensional (2D) plots in Figs. \ref{dnnfig3}(a)-\ref{dnnfig3}(c) for the same $k$ values which we have considered in 3D plots. In the 2D plots, one can clearly visualize  the changes that occur in frequency and amplitude of the $dn$- periodic wave around the RW. The amplitude of the RWs decreases as seen in Fig. \ref{dnnfig2} and also in Fig. \ref{dnnfig3}.  From this outcome, we conclude that when we increase the $k$ value the amplitude of the RW decreases.

In Fig. \ref{dnnfig5}, we show that RW arises on the top of $cn$-periodic wave background using Eq. (\ref{e38}) and (\ref{e8}) with (\ref{e13}) for three different values of $k$.  For $k=0.5$, RW on the cn-periodic background is shown in Fig. \ref{dnnfig5}(a) and its corresponding contour plot is displayed in Fig. \ref{dnnfig5}(b). At $|\hat{r}|=1.0$ the highest amplitude of RW is realized at origin $(x_0,t_0)$. While varying the elliptic modulus ($k$) from a lower to higher value, the amplitude of the RW increases. The amplitude of the RW is different for each $k$ value and they are found to be $|\hat{r}|=1.5~ (k=0.75)$ and  $1.8~(k=0.9)$. This observation is shown in Figs. \ref{dnnfig5}(c) and \ref{dnnfig5}(e) and their corresponding contour plots are displayed in Figs. \ref{dnnfig5}(d) and \ref{dnnfig5}(f), respectively.  We also observe that the frequency of periodic background wave changes in the $(x-t)$ plane. The 2D plots of the solution with the same $k$ values are presented in Figs. \ref{dnnfig4}(a)-\ref{dnnfig4}(c). Here also we observe that the frequency and amplitude of the $cn$- periodic wave around the RW varies. The amplitude of the RW increases as seen in Fig. \ref{dnnfig4} and also in Fig. \ref{dnnfig5}. More precisely, amplitude of the RW increases when we vary the $k$ value (from lower to higher).

\section{Conclusion}
In this work, with the help of one-fold DT formula we have obtained RW solution for the cmKdV equation on the elliptic function background.  The  admissible eigenvalues and the corresponding eigenfunctions are found through another procedure, namely nonlinearization of Lax pair. Using these expressions, we have generated the elliptic function background.  We then derived a second linearly independent solution (in the form of non-periodic) of the spectral problem with the same eigenvalues which in turn come out as a RW. We have analyzed the RWs on the periodic background for two different eigenvalues.  In the $dn$ case, our results reveal that while increasing the elliptic modulus value, the amplitude of RW on the periodic background decreases whereas the $cn$ case, amplitude of the RW increases. Our results may have potential applications in optics and oceanography.
%are potentially useful in optics experimentalists.
\section*{Acknowledgments}
NS thanks the University for providing University Research Fellowship.  KM wishes to thank the Council of Scientific and Industrial Research, Government of India, for providing the Research Associateship under the Grant No. 03/1397/17/EMR-II. The work of MS forms part of a research project sponsored by National Board for Higher Mathematics, Government of India, under the Grant No. 02011/20/2018NBHM(R.P)/R\&D 24II/15064.
%\section*{References}


\begin{thebibliography}{} 
	
	\bibitem{dudley}
	J. M. Dudley, G. Genty, A. Mussot, A. Chabchoub and F. Dias, Nat. Rev. Phys. \textbf{1}, 675-689 (2019).
	
	
	\bibitem{cha}
	A. Chabchoub, N. Hoffmann and N. Akhmediev, Phys. Rev. Lett. \textbf{106}, 204502 (2011).
	
	\bibitem{kib}
	B. Kibler, J. Fatome, C. Finot, G. Millot, F. Dias, G. Genty, N. Akhmediev and J. M. Dudley, Nat. Phys. \textbf{6}, 790-795 (2010).
	
	\bibitem{sch}
	J. A. Schuller, E. S. Barnard, W. Cai, Y. C. Jun, J. S. White, and M. L. Brongersma, Nat. Mater. \textbf{9}, 193 (2010).
	
	\bibitem{efi}
	V.B. Efimov, A.N. Ganshin, G.V. Kolmakov, P.V.E. McClintock and L.P. Mezhov-Deglin, Eur. Phys. J. Special Topics  \textbf{185}, 181–193 (2010).
	
	\bibitem{sha}
	M. Shats, H. Punzmann and H. Xia, Phys. Rev. Lett. \textbf{104}, 104503 (2010).
	
		
	\bibitem{kh}
	C. Kharif, E. Pelinovsky and A. Slunyaev, {\it Rogue Waves in the Ocean} (Springer, New York, 2009).
	
	\bibitem{wa}
	J. Wang, Q. W. Ma, S. Yan and A. Chabchoub, Phys. Rev. Appl. \textbf{9}, 014016 (2018).
	
	\bibitem{nak}
	N. Akhmediev, A. Ankiewicz and M. Taki, Phys. Lett. A \textbf{373}, 675-678 (2009).
	
	\bibitem{ak}
	N. Akhmediev, A. Ankiewicz and J. M. Soto-Crespo, Phys. Rev. E \textbf{80}, 026601 (2009).
	
	\bibitem{zhaqilao}	
	Zhaqilao, Phys. Scr. \textbf{87}, 065401 (2013).
	
	\bibitem{liu}	
	W. Liu and Y. Zhang, Phys. Scr. \textbf{95}, 045204  (2020).
	
	\bibitem{ref4}
	S. T. R. Rizvi, S. Ahmed, M. F. Nadeem and M. Awais, Optik \textbf{226}, 165955 (2021).
	
	\bibitem{ta}
    Y. Tao and J. He, Phys. Rev. E \textbf{85}, 026601 (2012).
    
    \bibitem{ya}
    X. W. Yan, S. F. Tian, M. J. Dong and L. Zou, Nonlinear Dyn. \textbf{92}, 709-720 (2018).
    
    \bibitem{oh}
    Y. Ohta , and J. Yang, Phys. Rev. E \textbf{86}, 036604 (2012).
    
    \bibitem{ref1}
    M. Younis, S. Ali, S. T. R. Rizvi, M. Tantawy, K. U. Tariq and A. Bekir, Commu. Non. Sci. Num. Sim. \textbf{94}, 105544 (2021).
    
    \bibitem{ref2}
    A. R. Seadawy, S. T. R. Rizvi, S. Ahmed, M. Younis and D. Baleanu, Open Phys. \textbf{19}, 1-10 (2021).
    
    \bibitem{ref3}
    S. T. R. Rizvi, A. R. Seadawy, F. Ashraf, M. Younis, H. Iqbal and D. Baleanu, Results in Phys. \textbf{19}, 103661 (2020).
    
    
     \bibitem{ref5}
    S. T. R. Rizvi, A. R. Seadawy, S. Ahmed, M. Younis and K. Ali, Int. J. Mod. Phys. B \textbf{35}, 2150055 (2021).
    
      \bibitem{wan}
    L. H. Wang, K. Porsezian and J. S. He, Phys. Rev. E \textbf{87}, 053202 (2013).
	
	  \bibitem{guo}
	B. Guo, L. Ling and Q. P. Liu, Phys. Rev. E \textbf{85}, 026607 (2012).
	
	  \bibitem{vi}
	N. Vishnu Priya , and M. Senthilvelan, Commu. Non. Sci. Num. Sim. \textbf{20}, 401-420 (2015).
	
	\bibitem{agaf}
	D. S. Agafontsev and V. E. Zakharov, Nonlinearity \textbf{28},  2791-2821 (2015).
	
	\bibitem{mu}
	G. Mu, Z. Qin and R. Grimshaw, SIAM J. Appl. Math. \textbf{75},  1-20 (2015).
	
	\bibitem{agaf1}
	D. S. Agafontsev, and V. E. Zakharov, Nonlinearity \textbf{29}, 3551-3578 (2016).
	
	\bibitem{mani}
	K. Manikandan, P. Muruganandam, M. Senthilvelan and M. Lakshmanan, Phys Rev E.  \textbf{90},  062905 (2014).
	
	\bibitem{mani2}
	K. Manikandan, P. Muruganandam, M. Senthilvelan and M. Lakshmanan, Phys. Rev. E  \textbf{93}, 032212 (2016).
	
	\bibitem{xue}
	B. Xue, J. Shen and X. Geng, Phys. Scr. \textbf{95}, 055216 (2020).
	
	\bibitem{kedziora}
	D. J. Kedziora, A. Ankiewicz and N. Akhmediev, Euro. Phys. J. Spec. Topics \textbf{223}, 43-62 (2014).
	
	\bibitem{chen2}
	J. Chen and D. E. Pelinovsky, Proc. R. Soc. A \textbf{474}, 20170814 (2018). 
	
	\bibitem{chen1}
	J. Chen and D. E. Pelinovsky, Nonlinearity \textbf{31}, 1955 (2018).
	
	\bibitem{chen3}
	J. Chen and D. E. Pelinovsky, J. Nonlinear Sci. \textbf{29}, 2797 (2019). 
	
	\bibitem{peng}
	W. Q. Peng, S. F. Tian, X. B. Wang and T. T. Zhang, Wave Motion \textbf{93}, 102454 (2020).
	
	\bibitem{pel}
D. E. Pelinovsky and R. E. White, Proc. R. Soc. A \textbf{476}, 2242 (2020).
	
	
	\bibitem{sinthu1}
	N. Sinthuja, K. Manikandan and M. Senthilvelan, Formation of rogue waves on the periodic background in a fifth-order nonlinear Schr\"odinger equation, 2021, arXiv:2102.11523.
	
	\bibitem{n1}
	V. Aleshkevich, Y. Kartashov, V. Vysloukh, Opt. Comm. \textbf{185}, 305 (2000).

	\bibitem{n3}
	P. Leboeuf, N. Pavloff, Phys. Rev. A \textbf{64}, 033602 (2001).
	
	\bibitem{n4}
	Y. V. Bludov, V. Konotop, N. Akhmediev, Eur. Phys. J. Special Topics \textbf{185}, 169 (2010).
	
	\bibitem{chen5}
	J. Chen, D. E. Pelinovsky, and R. E. White, Phys. Rev. E \textbf{100}, 052219 (2019). 
	
	\bibitem{sinthu}
	N. Sinthuja, K. Manikandan and M. Senthilvelan, Euro. Phys. J.  Plus  \textbf{136}, 305 (2021).
	
	\bibitem{zhou}
	R. G. Zhou, J. Math. Phys.  \textbf{48}, 013510 (2007).
	
	\bibitem{zhou1}
	R. G. Zhou, Stud. Appl. Math. \textbf{123}, 311 (2009).
	
	\bibitem{crabb}
	M. Crabb and N. Akhmediev, Phys. Rev. E \textbf{99}, 052217 (2019) .
	
	
	\bibitem{zhang1}
	Z. Zhang, X. Yang and B. Li, Nonlinear Dyn. \textbf{100}, 1551-1557 (2020).
	
	\bibitem{he}
	J. He, L. Wang, L. Li, K. Porsezian and R. Erdelyi, Phys. Rev. E  \textbf{89}, 062917 (2014).
	
	
	\bibitem{zhan}
	Z. Zhang, X. Yang and B. Li, Appl. Math. Lett. \textbf{103}, 106168 (2020). 
	
	\bibitem{yuan}
	L. Y. Ma, S. F. Shen and Z. N. Zhu, J. Math. Phys. \textbf{58}, 103501 (2017). 
	
\end{thebibliography}
\end{document}